\documentclass[12pt]{article}
\usepackage[margin=1in]{geometry}
\usepackage{amsmath,amssymb,amsthm,mathrsfs}
\usepackage{hyperref}
\usepackage{apacite}
\usepackage{natbib} 
\usepackage{graphicx}
\usepackage{caption}
\usepackage{enumitem}
\usepackage{mathtools}
\usepackage{tikz}
\usepackage[protrusion=true, expansion=false]{microtype}

\newcommand{\be}{\begin{equation}}
\newcommand{\ee}{\end{equation}}

\newcommand{\G}{\mathcal{G}}

\newtheorem{defi}{Definition}

\theoremstyle{remark}

\title{Making Symmetry Explicit:\\ \Large{The Limits of Sophistication}.}
\author{Henrique de Andrade Gomes}
\date{\today}

\begin{document}
\maketitle

\begin{abstract}

Symmetry is often treated in philosophy of physics as an interpretive problem. A particularly lively dispute concerns \emph{local} symmetries: do they indicate surplus structure that ought to be expunged, or are they merely a harmless redundancy? One influential response favours the second option for certain theories---those dubbed \emph{internally sophisticated}. And indeed, in much of physics practice, local symmetries are left implicit: one simply works ``up to isomorphism'' without pausing over invariance. But not always. In some settings, local symmetry and invariance become pressing practical concerns for physicists. Yet philosophical discussions of sophistication have paid little sustained attention to when, and why, this happens.

Surveying textbook general relativity (GR) and gauge theory, I identify the \emph{settings} in which diffeomorphism invariance or gauge invariance must be handled explicitly. (Here a \emph{setting} is a choice of representational framework or background assumptions within which one formulates and uses the theory---for instance, linearisation, an initial-value formulation, or a Hamiltonian $3+1$ formalism.) I propose an operational criterion---\emph{background-relative sophistication} (BRS)---and argue that it accounts well for the pattern: it marks just where symmetry can stay implicit and where it must be made explicit. Quantum and subsystem settings raise a further difficulty: there, certain \emph{tasks} (superposition and gluing) force symmetry into view even for theories that are BRS.

\end{abstract}

\tableofcontents

\section{Introduction}\label{sec:intro}

When, if ever, is symmetry in a physical theory conceptually problematic? A large literature asks when symmetry-related models represent genuinely distinct possibilities, and what to do when they don't.\footnote{To keep the discussion anchored without attempting an exhaustive bibliography, I will treat the following as touchstones. For Leibniz equivalence and the hole argument: \citet{EarmanNorton1987, Weatherall_hole, Pooley_Read}. For gauge-theoretic analogues and the connection to surplus structure: \citet{Belot2003, Belot_sym}. For sophistication (and its critics): \citet{Dewar2017, ReadMartens, MarchRead2025, BradleyWeatherall_soph}.} In general relativity, diffeomorphism-related models are standardly taken to represent the same physical possibility (Leibniz equivalence); in gauge theory, the same goes for gauge-related connections or potentials. But what, if anything, should one \emph{do} about this redundancy?

One influential answer is: nothing much. \textit{Internal sophistication} holds that, since mathematical objects are determined only up to isomorphism, the relevant isomorphisms already preserve representational content; thus if symmetries are isomorphisms they need not be eliminated from the formalism. And much of physics practice seems to agree. Canonical presentations of GR do not treat local symmetries as standing interpretive puzzles. Textbook presentations do not pause before writing the geodesic equation or the Einstein field equations to announce that these objects are defined only up to isomorphism; much less do they seek a formalism whose models represent physical possibilities one-to-one. Rather, they work in a representational language---tensors on a manifold, connections on a principal bundle---whose grammar is already adapted to the relevant equivalence relation. If sophistication were the whole story, symmetry would be invisible in routine physics.

But it is not invisible everywhere. There are \emph{settings} in which symmetries \emph{must} be handled explicitly---settings in which the standard presentations introduce gauge-fixing, constraint analysis, dressings, or other overt machinery. Belot and Earman have pressed this point, arguing that the hole argument is not merely a philosophical curiosity but reflects something physicists themselves recognise as a live issue:
\begin{quote}
Far from dismissing the hole argument as a simple-minded mistake which is irrelevant to understanding GR, many physicists see it as providing crucial insight into the physical content of GR. \citep[p.~169]{belot_earman_1999}
\end{quote}
So we have a mixed practice: symmetry is sometimes invisible, sometimes unavoidable. This paper asks what explains the difference.

Of course, philosophy of physics should not simply defer to practice. Even if textbooks leave symmetry implicit, one can still ask what the theory's models are, what fixes the referents of their mathematical structures, and what counts as sameness of physical situations. It is one thing to write $G_{ab}=8\pi T_{ab}$ as a tensor equation; it is another to say what fixes the referent of $G_{ab}$. Nonetheless, if metaphysics is to be responsive to our best theories, it should be attentive to the way those theories are actually used---and in particular, to the pattern of when they do and do not make symmetry explicit.

\subsection*{Plan of the paper}
\label{sec:plan}

Throughout, I say that symmetry becomes \emph{explicit} when overt symmetry-handling machinery---gauge-fixing, constraints, dressings, etc---is introduced.

The pattern that emerges from the case studies is fairly sharp:
\begin{itemize}
\item In classical GR, diffeomorphism invariance becomes explicit in essentially three settings: (i) linearisation, (ii) the initial value problem, and (iii) the $3+1$ (ADM) Hamiltonian formalism.

\item In gauge theory, gauge invariance is explicit in a potential-based setting but largely remains implicit in a principal-bundle setting.
\item In both theories, symmetry comes to the fore again in two further settings: quantisation and the treatment of regional subsystems.
\end{itemize}

Section~\ref{sec:sophistication} introduces the operational criterion---\emph{background-relative sophistication} (BRS): a theory is BRS for a symmetry group $S$ when $S$ acts as automorphisms of the admissible background structure the formalism already holds fixed. 

The empirical question then is: are the settings where BRS fails the same settings where symmetry is made explicit in canonical presentations?\footnote{For the purposes of this paper I adopt BRS only instrumentally. I treat it as a hypothesis to be tested against textbook practice, not as the final word on what internal sophistication means.} Section~\ref{sec:GR} argues that for GR they are: each of the three GR settings introduces background structure that generic diffeomorphisms do not preserve. Section~\ref{sec:gauge} makes the same point for gauge theory: in the potential formalism, gauge transformations act affinely on local representatives and must be explicitly managed; in the principal-bundle formalism, they are vertical bundle automorphisms---automorphisms of the background itself---and can remain implicit. Section~\ref{sec:quantum} takes up a further difficulty. Even for theories that satisfy BRS, the construction of local observables, superposition, and the gluing of regional descriptions all require cross-model or cross-region comparison that forces symmetry-handling machinery back into view.   The upshot of these Sections is that symmetry can stay implicit when BRS holds and the work at hand is exhausted by single-model description; when either condition fails, symmetry must be made explicit.  Section~\ref{sec:conclusion} draws the moral.

\section{Background-relative sophistication}
\label{sec:sophistication}

If sophistication is to do more than restate the slogan that 
symmetry-related models represent the same possibility, it 
needs a criterion sharp enough to predict where symmetry can be omitted and where it cannot. This section supplies one. It also supplies the methodology by which to compare these predictions to practice, a topic to which I now turn.

\subsection{Surveying the literature}\label{sec:survey}
I will say canonical presentations of a theory treat a symmetry as `unproblematic' in a given setting simply if questions about that symmetry do not arise as \emph{standing representational or interpretive problems}. The survey rests on four checks, 
applied to standard expository sources:
\begin{enumerate}[label=(\alph*)]
\item \textbf{Index and keyword checks.} Whether the symmetry recurs in the index and chapter headings (e.g.\ ``diffeomorphism invariance'', ``general covariance'', ``gauge invariance'', ``gauge fixing'', ``relational", ``observable"). 
\item \textbf{Derivation checks.} Whether core derivations invoke the symmetry as a nontrivial step (e.g.\ by showing how quantities transform, or by asserting invariance) rather than simply writing tensorial or bundle-covariant expressions.
\item \textbf{Machinery checks.} Whether the presentation introduces any of the characteristic overt devices: gauge-fixing conditions, constraint analysis and constraint algebras, the introduction of reduced variables, explicit quotient constructions, physical reference frames, dressings, or boundary/edge-mode extensions.
\item \textbf{Failure-mode checks.} Whether the symmetry is invoked to diagnose a breakdown of na\"ive reasoning (e.g.\ non-uniqueness in the initial value problem, or spurious degrees of freedom in linearisation).
\end{enumerate}

When a symmetry can remain implicit, standard expositions tend to proceed in a formal language already adapted to it (tensor calculus on a manifold, or bundle-covariant objects), without introducing extra representational structure. When the task instead requires one to keep track of symmetry explicitly---for example, to count degrees of freedom, to specify initial data, or to compare models or subsystems---expositions introduce the relevant symmetry-handling tools. The difference is not subtle.

I applied these checks to a small set of widely used textbooks and lecture-note sources, aiming for ``standard'' expository treatments rather than specialist research monographs. For classical GR, the core sources were \citep{MTW}; \citep{Wald_book}; \citep{Carroll2019}; \citep{Poisson_book}; and, for the $3+1$ and initial-value formalisms, more specialised sources such as \citep{3+1_book}. For gauge theory, I contrasted two families of presentations: potential-based and principal-bundle based. As representatives of the potential tradition I used standard electrodynamics and field-theory texts that treat gauge freedom via potentials, gauge conditions, and constraints, such as \citep{aitchison2012gauge, peskin1995introduction, cheng1984gauge,schwartz2014quantum}. As representatives of the bundle tradition I used standard geometry-and-physics texts that build gauge covariance into the basic objects, such as \citep{Bleecker}; \citep{kobayashivol1}; \cite{Nakahara}; \citep{Baez_book}. 

The upshot is that how prominently local symmetry figures depends both on what background is assumed and on the task at hand. The same theory can be presented in a way that makes symmetry practically invisible in routine calculation, yet forces it into view in regimes where a different notion of background is employed or where one needs to compare data in a way that the background grammar does not automatically control.

\subsection{A definition of BRS}
\label{subsec:definition}

Can sophistication's deflationary attitude be sharpened into a criterion that predicts where symmetry becomes explicit?

 To make the question precise, it helps to set sophistication against the alternative: \emph{reduction}. A reduced theory traffics only in symmetry-invariant quantities, eliminating any reference to the non-invariant structure. Reduction and sophistication agree that symmetry-related models represent the same physical situation, but they implement that agreement differently: reduction removes the redundant structure from the formalism, whereas sophistication retains the standard variables while treating symmetry-related models as (for representational purposes) `notational variants'. In general relativity, a reduced formulation might describe spacetime geometry purely in terms of integrals of curvature invariants or abstract geodesic structure, objects without explicit dependence on manifold points. In gauge theory, a reduced formulation might describe the physics purely in terms of holonomies, without reference to connections or gauge potentials. When canonical presentations aim for such reduction, that is itself a sign that the symmetry is being treated as problematic. The contrast matters in practice, since reduced formalisms can be technically cumbersome or unavailable, whereas sophistication preserves the familiar calculational apparatus.

The familiar slogan for 
sophistication---that when symmetry-related models are 
isomorphic they represent the same possibility---is not 
sharp enough to serve as a criterion. It does not say what, 
in a given setting, is to count as the relevant notion of 
isomorphism. 

Assume then, that the models of a theory $T$ are given as ordered pairs
\be
(B, \varphi), \qquad \varphi \in \Gamma(E),
\ee
where $B$ is the \emph{background structure} (e.g.\ the smooth structure of a manifold $M$; the linear structure of a vector bundle $E \to M$; a chosen foliation; a fixed background metric; a boundary structure; or any other structure treated as fixed across the class of models under consideration) and $\varphi$ is the \emph{dynamical content} (metric fields, connections, matter fields, momenta; or any other content that is treated as variable across the class of models under consideration).  The point of separating $B$ from $\varphi$ is not to settle a substantive background/dynamics dispute, but to expose a parameter of variation in representational practice: the same physical theory can be cast in formalisms with different choices of what is treated as background.

Let $\mathrm{Aut}(B)$ be the group of automorphisms of the background structure $B$. This group acts on $\varphi$ by pullback (or pushforward, as appropriate), yielding an induced action $\varphi \mapsto \varphi^f$ for $f \in \mathrm{Aut}(B)$. Let $S$ be a group of dynamical symmetries of $T$: for example, a variational symmetry group, or a group of transformations that map solutions to solutions.\footnote{Again, for our purposes, nothing important hangs on precisely this group is to be defined. Suffice it to say that each theory described here in practice has a well-defined dynamical symmetry group.} This $S$ may coincide with $\mathrm{Aut}(B)$, it may be larger, or it may stand in a more complicated relationship to the background.

The central definition is then the following.
\begin{defi}[\textbf{Background-Relative Sophistication (BRS)}] A theory $T$ is \emph{background-relative sophisticated for $S$} (BRS) if and only if $S \subseteq \mathrm{Aut}(B)$, where $B$ is the admissible background structure for the setting under consideration.\footnote{This definition is close in spirit to Earman's symmetry principles SP1 and SP2 \cite{Earman_world}[p. 45]. A structurally similar characterisation appears in \cite{Jacobs_thesis}[p. 101], where he glosses the relevant notion as the ``stipulation of a set of relations'' on sub-domains that symmetry-inducing bijections leave invariant.}\label{defi:soph}
\end{defi}

A worry immediately arises: if $B$ were a free choice, one could trivialise the criterion by simply building the desired symmetries into $B$ by fiat. So if the criterion is to have any bite, we need a constraint on what counts as an \emph{admissible} background. I take this to be practice-sensitive: $B$ is whatever the formulation in the setting under consideration \emph{actually holds fixed} (and needs held fixed to carry out its characteristic tasks). In particular, $B$ must be fixed across the class of models in that setting, and it must be fixed \emph{by the formulation itself} (not merely by some alternative reformulation one might prefer). The question, then, is not whether one \emph{could} choose a background relative to which a symmetry becomes an automorphism, but whether the background that physics \emph{in fact introduces} in that domain has that property.
 The constraint follows from what `in a given setting' already means: the setting fixes the representational apparatus; we are not free to swap it out.

When $T$ is BRS, $S$-related models arguably differ only by a relabelling that the background structure treats as insubstantial; they differ only in which bare points---of \(M\), of $P$, or of a vector bundle $E$---are assigned which structural roles, while the overall structural profile stays the same.\footnote{See \citep{Gomes_intgeom} for a `geometry-first' description of gauge theory solely in terms of vector bundles.\label{ftnt:geom}} In the spacetime case, for instance, two models related by a diffeomorphism agree on the \emph{same geometric structure} but disagree about which manifold point plays which part.\footnote{The remaining difference between the models is exactly the kind of difference that anti-haecceitism denies is physically real. In the context of spacetime, anti-haecceitism is the view that spacetime points have no primitive identity over and above the qualitative and structural roles they play. In general, the slogan characterisation is: ‘qualitative properties and relations are those that do not invoke particular objects in their definitions’. They can be illustrated
with paradigmatic examples: ‘loves’, ‘is taller than’ are qualitative; ‘loves Juliet’, ‘is taller than Romeo’, are not.\label{ftnt:anti}}

For illustration, take an ordinary globe: a round sphere with a fixed distribution of landmasses  representing Earth. Let the background $B$ be the round metric on $S^2$, so $\mathrm{Aut}(B)\cong SO(3)$. The landmass  distribution is a field $\varphi$ on this background (e.g.\ a characteristic function of `land' versus `sea'), and it is not rotation-invariant.

Nonetheless there is no interpretive difficulty: rotating the globe does not produce a distinct geography; it is the same object described by a different assignment of points of $S^2$ in our representation. The representation need not be invariant because the background already supplies the relevant standard of sameness: fields related by an automorphism of $B$ represent the same globe. Nobody worries about the `gauge freedom' of the globe.



\paragraph{Summary.}
The expectation is this. When the admissible background 
already admits $S$ as automorphisms, symmetry can remain 
implicit: one works in an $S$-covariant language and ignores the symmetries. When the background does not admit $S$, 
symmetry must be made explicit---via gauge-fixing, 
constraint analysis, dressings, or some other device---
because the formalism no longer internalises it. 
The next two sections test this expectation against 
classical GR and gauge theory.

\section{When spacetime diffeomorphisms become explicit}
\label{sec:GR}

Classical general relativity forces diffeomorphism invariance into explicit view in a surprisingly narrow and consistent set of settings---precisely those in which BRS fails.

Consider two models \((M,g)\) and \((M,\phi^*g)\) of general relativity, related by a diffeomorphism \(\phi:M\to M\) (an automorphism of the smooth background structure of $M$, held fixed across models).\footnote{One could add that the fact that we are dealing with metrics of a certain signature is also held fixed across models, and so `being a Lorentzian metric' is part of `the background'. The important point here is that the intersection of the automorphisms of all those structures would still be $\text{Diff}(M)$.}  By Definition~\ref{defi:soph}, this case counts as BRS.

And when we compare this expectation with canonical presentations, the fit is good. According to the survey described in Section \ref{sec:survey}, diffeomorphism invariance is rarely discussed at length in the covariant formulation. Wald's textbook is representative; the main idea is only described in an appendix:
\begin{quote}
    If a theory describes nature in terms of a spacetime manifold, $M$, and tensor fields, $T^{(i)}$, defined on the manifold, then if $\phi: M \to N$ is a diffeomorphism, the solutions $(M, T^{(i)})$ and $(N, \phi^* T^{(i)})$ have physically identical properties. Any physically meaningful statement about $(M, T^{(i)})$ will hold with equal validity for $(N, \phi^* T^{(i)})$. \ldots\ Thus, the diffeomorphisms comprise the gauge freedom of any theory formulated in terms of tensor fields on a spacetime manifold' \citep[p.~438]{Wald_book}.
\end{quote}

For example, the Einstein field equation $G_{ab}[g]=8\pi T_{ab}[g,\psi]$ is written as a tensor equation on a manifold. If \(\phi:M\to M\) is a diffeomorphism, then pulling back the fields yields another model \((M,\phi^*g,\phi^*\psi)\), and both sides of the equation pull back together: $G_{ab}[\phi^*g]=\phi^*G_{ab}[g]$ and $T_{ab}[\phi^*g,\phi^*\psi]=\phi^*T_{ab}[g,\psi]$. The tensorial character of the two equations guarantees that one is satisfied if and only if the other one is. Once one works with covariant objects, invariance is built into the grammar of the claims rather than enforced by an extra step. So much for covariant GR.

But this is not always the case. \cite{HawkingEllis} perfectly encapsulate the standard practice, while forewarning the reader about a setting in which symmetry must be explicitly invoked:
\begin{quote}
    Two models \((\mathcal{M},g)\) and \((\mathcal{M}',g')\) will be taken to be equivalent if they are
\emph{isometric}, that is if there is a diffeomorphism \(\theta:\mathcal{M}\to\mathcal{M}'\) which
carries the metric \(g\) into the metric \(g'\), i.e.\ \(\theta_{*}g=g'\). Strictly speaking then,
the model for space-time is not just one pair \((\mathcal{M},g)\), but a whole equivalence class
of all pairs \((\mathcal{M}',g')\) which are equivalent to \((\mathcal{M},g)\). We shall normally
work with just one representative member \((\mathcal{M},g)\) of the equivalence class, but \textit{the fact that this pair is defined only up to equivalence is important in some situations, in particular in the discussion of the Cauchy problem} in chapter~7. (p. 56, my italics)
\end{quote}

Indeed, standard textbooks on classical general relativity do sometimes foreground diffeomorphism invariance, but in a remarkably consistent pattern: across the mainstream literature, it becomes explicit in essentially three settings.\par
\begin{enumerate}
\item linearisation (degree-of-freedom counting, gauge-fixing, and propagators),
\item the initial value problem (constraints, well-posedness, and determinism),
\item the $3+1$ (ADM) formalism (foliation dependence, constraint generators, and the status of `gauge').
\end{enumerate}
This pattern matches what we would expect from BRS, as I will now argue. 

\subsection{Linearised gravity }
\label{subsec:linearised}

The first setting, a context or sector of a theory in which the failure of BRS occurs, is linearised gravity. Here, textbooks routinely discuss gauge transformations, gauge-fixing conditions, and gauge-invariant degree-of-freedom counting, via: helicity decomposition, transverse-traceless gauge, harmonic gauge, and so on \citep{MTW, Poisson_book, Wald_book}. The reason is not far to seek. The linearised formalism treats the metric perturbation $h_{\mu\nu}$ as a field on a fixed background $\eta_{\mu\nu}$:
\be
g_{\mu\nu} = \eta_{\mu\nu} + h_{\mu\nu}, \qquad |h_{\mu\nu}| \ll 1.
\ee
This changes the original background. The representational background $B'$ now includes the flat metric $\eta$ (or at least a preferred flat structure), and the automorphisms of that background are correspondingly restricted (to isometries of $\eta$, or diffeomorphisms preserving the chosen background structure). But the linearised equations still admit a gauge freedom parametrised by an arbitrary vector field $\xi$: 
\be
h_{\mu\nu} \mapsto h_{\mu\nu} + \partial_\mu \xi_\nu + \partial_\nu \xi_\mu.
\ee

In the language of Section~\ref{sec:sophistication}, linearisation leads to failure of BRS:
\be
S = \mathrm{diff}(M) \not\subseteq \mathrm{Aut}(B'), \qquad B' = (M, \eta).
\ee
Once a rigid background is installed, one is no longer studying geometry in the covariant sense; one is studying field theory on a fixed stage. The transformation $h \mapsto h + \mathcal{L}_\xi \eta$ is not the action of an automorphism of $B'$; it is a transformation of the dynamical field that leaves the linearised equations invariant. In that setting, it becomes possible (indeed easy) to write down expressions that appear to carry physical content but in fact do not.

The textbook emphasis on gauge is then forced for most practical questions. Two familiar examples are: (i) \emph{degree-of-freedom counting}: extracting the wave equation, or the two radiative (gravitational-wave) degrees of freedom requires either gauge-invariant constructions or a gauge choice (e.g.\ harmonic and transverse--traceless gauge, respectively); and (ii) \emph{inversion and propagators}: the quadratic form in the perturbative action has gauge zero-modes, so inversion (and hence propagator definition) requires gauge-fixing. Similar remarks apply in applications such as the direct Newtonian limit (i.e. not going through Newton-Cartan; cf \cite{Wald_book}[Ch. 3.4]).

In sum, linearised presentations make gauge explicit because the approximation introduces a representational arena in which the symmetry no longer comes from the background.

\subsection{The initial value problem and the \texorpdfstring{$3+1$}{3+1} formalism}
\label{subsec:ivp}

The second and third contexts are best treated together: both recast general relativity in terms of data on spacelike hypersurfaces, and thereby transform diffeomorphism freedom into explicit constraints and `gauge' freedom.

To pose an initial value problem one chooses a spacelike hypersurface $\Sigma$ and specifies Cauchy data. In vacuum general relativity these data may be taken as a Riemannian metric $\gamma_{ij}$ on $\Sigma$ together with its extrinsic curvature $K_{ij}$ (or equivalently the canonical pair $(\gamma_{ij},\pi^{ij})$). Not every pair is admissible: the data must satisfy the constraint equations:\footnote{These correspond to the time-time and time-space components of the Einstein equations. If the dynamical equations involved higher order tensors, more constraints would be implied by the Gauss-Codazzi-Ricci embedding equations.}
\begin{align}
R[\gamma] - K_{ij}K^{ij} + K^2 &= 0 \qquad \text{(Hamiltonian constraint)},\label{eq:G00} \\
D_j(K^{ij} - \gamma^{ij} K) &= 0 \qquad \text{(momentum constraints)}\label{eq:G0i},
\end{align}
where $R[\gamma]$ is the scalar curvature of $\gamma$ and $D$ is its Levi--Civita derivative. A precise well-posedness theorem then says (roughly) that, for suitable data on the constraint surface, there exists a spacetime development and it is unique only \emph{up to} diffeomorphism. In other words, diffeomorphism freedom enters twice: it identifies distinct representatives of the same physical Cauchy data, and it is built into the very sense in which solutions are unique \citep{choquet2008, ChoquetBruhat1969}.

The ADM $3+1$ formalism extends the choice of $\Sigma$ in the IVP with an explicit foliation and a Hamiltonian re-expression of the dynamics. One assumes a splitting $M\cong\mathbb{R}\times\Sigma$ and introduces lapse $N$ and shift $N^i$, writing the metric as
\be
ds^2 = -N^2 dt^2 + \gamma_{ij}(dx^i + N^i dt)(dx^j + N^j dt).
\ee

The configuration space is then the space of Riemannian metrics on a fixed three-manifold $\Sigma$, with momenta related to $K_{ij}$; the Einstein equations decompose into evolution equations plus the constraints \eqref{eq:G00} and \eqref{eq:G0i} re-written in terms of the canonical momenta, which can now be understood as generators of gauge (via symplectic flow).

\begin{figure}[t]
\centering
\includegraphics[width=0.6\textwidth]{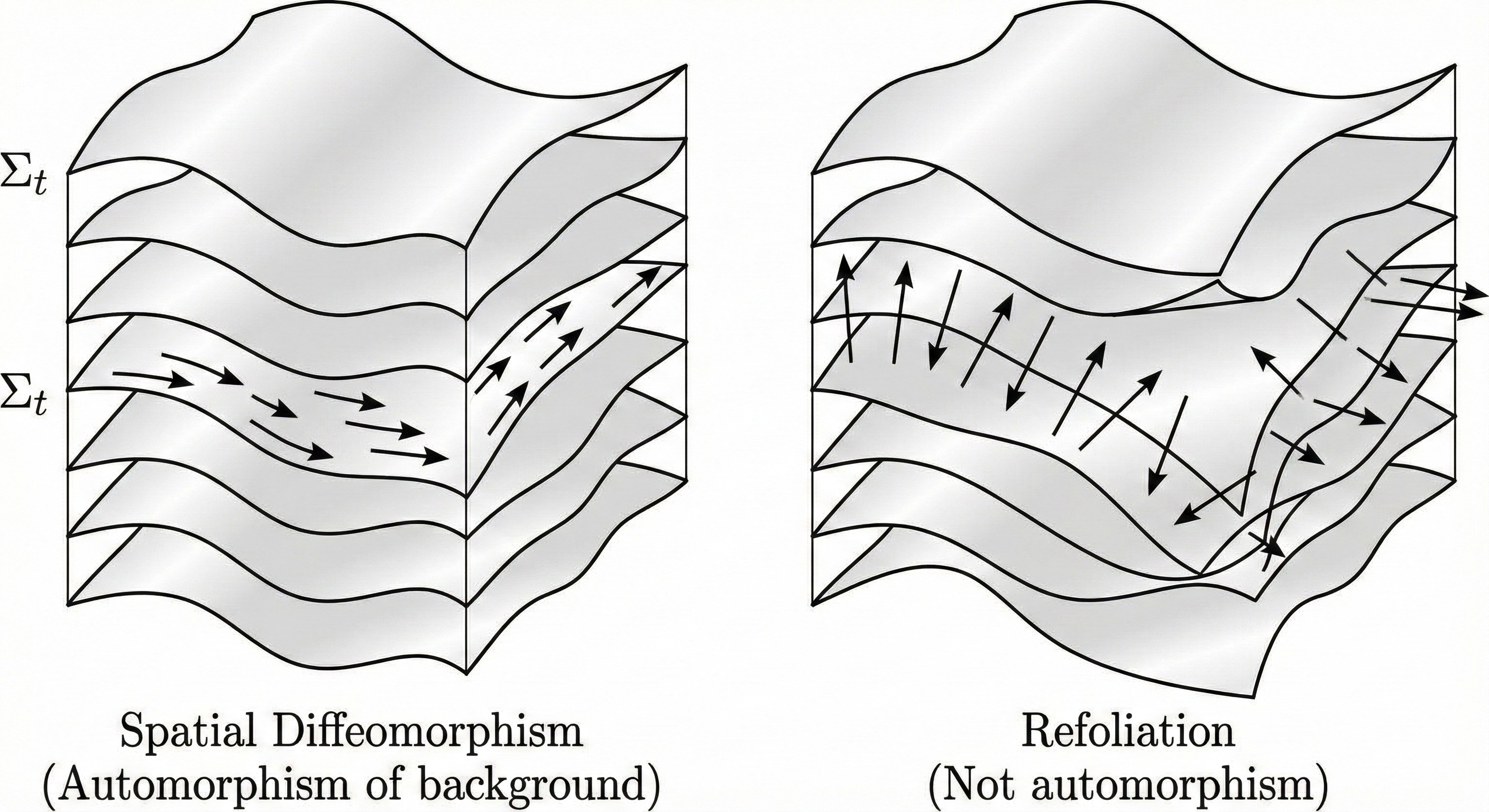}
\caption[Spatial diffeomorphisms vs.\ refoliations]{Left: a spatial diffeomorphism slides points along the leaves $\Sigma_t$ of a foliation, preserving the foliation structure; it is an automorphism of the ADM background. Right: a refoliation deforms the hypersurfaces in the normal direction; the resulting transformation depends on the dynamical fields and does not preserve the foliation structure. BRS holds for the former but fails for the latter.}
\label{fig:diffs_refol}
\end{figure}

Note that now generic diffeomorphisms do not preserve the chosen foliation: once the foliation data are treated as part of the admissible background $B_{\mathrm{ADM}}$, the full dynamical symmetry group corresponding to the diffeomorphisms is not realised as automorphisms of that background,
\be
\mathrm{Diff}(M) \not\subseteq \mathrm{Aut}(B_{\mathrm{ADM}}).
\ee
Relatedly, the constraint structure itself splits (see Figure~\ref{fig:diffs_refol}). The momentum constraints generate spatial diffeomorphisms within a leaf $\Sigma_t$, which \emph{are} automorphisms of the spatial background and so can be absorbed in the BRS sense.

This can be seen directly in the phase-space action of the momentum constraint. In ADM variables, the momentum constraint is
\be
\mathcal{H}_i \;=\; -2\,\gamma_{ij}D_k\pi^{jk},
\ee
and for any vector field $v^i$ on $\Sigma$ one considers the smeared generator
\be
D(v)\;:=\;\int_{\Sigma} d^3x\; v^i\,\mathcal{H}_i.
\ee
Its action on the canonical data is computed by the Poisson bracket. One finds
\be
\{\gamma_{ij},D(v)\}=\mathcal{L}_v\gamma_{ij},
\qquad
\{\pi^{ij},D(v)\}=\mathcal{L}_v\pi^{ij},
\ee
so for any functional $F[\gamma,\pi]$ on phase space,
\be
\{F,D(v)\}=\delta_v F,
\ee
i.e. the momentum constraint generates the infinitesimal spatial diffeomorphism along $v$.

The Hamiltonian constraint instead generates refoliations---normal deformations of $\Sigma$---and these do not assemble into a Lie group action on a single fixed background. This is reflected in the hypersurface-deformation algebra: the commutator of two Hamiltonian constraints involves the dynamical metric,
\be
\{H(N), H(M)\} = \;\int_{\Sigma} d^3x\; \mathcal{H}_i(\gamma^{ij}(N \partial_j M - M \partial_j N)),
\ee
so the ``structure constants'' are in fact field-dependent structure functions. What this field-dependence tells us is that refoliations are not implemented as background automorphisms shared across the whole phase space of $(\gamma,\pi)$---they depend on the particular model through the metric.
I think this is the single most telling illustration of the criterion. Spatial diffeomorphisms and refoliations are both `gauge' in the loose sense, yet they behave quite differently by the lights of BRS---and this difference is exactly what the constraint algebra encodes.

As \citet{ThebaultGryb_hole} emphasise, this also marks a mismatch between the canonical and covariant pictures: the canonical transformations generated by the Hamiltonian constraint should not be straightforwardly identified with the covariant action of $\mathrm{Diff}(M)$ on four-dimensional fields.

In sum, BRS fails for this part of the symmetry. As an aside, this is one reason the Hamiltonian formulation is the natural home of the `problem of time' in canonical quantum gravity \citep{Isham_POT, Kuchar_Time}: the Hamiltonian is (on shell) a sum of constraints, so its vanishing can be read as a challenge to straightforward notions of evolution. Whatever one makes of that diagnosis, it reflects the same underlying point: the Hamiltonian constraint functions as a nontrivial constraint rather than as a redundancy that can be absorbed into the identity conditions for models.

\paragraph{Summary (GR).}
In covariant classical GR, diffeomorphism invariance can 
remain largely implicit: tensor equations on a manifold are 
naturally read up to isomorphism. It becomes explicit in 
just three settings---linearisation 
(rigid background~$\eta$), the initial value problem 
(constraints and uniqueness), and the $3+1$ formalism 
(foliation dependence and the field-dependent 
hypersurface-deformation algebra)---and in each case the 
reason is the same: the setting introduces background 
structure that diffeomorphisms do not preserve.

\section{Gauge symmetry across formalisms}\label{sec:gauge}

Gauge theory provides a particularly clean test of BRS, because the same physics is routinely presented in two quite different but equivalent formalisms---a principal-bundle formalism (Subsection~\ref{subsec:PFB}) and a gauge-potential formalism (Subsection~\ref{subsec:gaugepotentials})---and the contrast shifts where the gauge freedom lives.

\subsection{Principal bundles and background-relative sophistication}\label{subsec:PFB}

The principal-bundle formalism is designed so that gauge transformations are literally automorphisms of the admissible background structure. In that sense, it builds gauge freedom into the background notion of sameness, and so (by the criterion of Definition~\ref{defi:soph}) realises BRS.

The primitive object is a principal \(G\)-bundle \(\pi:P\to M\) where \(P\) is a smooth manifold equipped with a free and transitive \emph{right} action \(R_g:P\to P\), \(p\mapsto p\cdot g\), such that the orbit space \(P/G\) is (smoothly) identified with the base manifold \(M\). For present purposes the admissible background structure \(B\) includes at least the bundle \(\pi:P\to M\) together with its \(G\)-orbit structure. The corresponding notion of background automorphism is a \(G\)-equivariant bundle automorphism \(\Phi:P\to P\), i.e.\ a diffeomorphism satisfying \(\Phi(p\cdot g)=\Phi(p)\cdot g\) and covering some diffeomorphism \(f:M\to M\) in the sense that \(\pi\circ\Phi=f\circ\pi\). The \emph{vertical} automorphisms, with \(f=\mathrm{id}_M\), are the gauge transformations.

The basic dynamical object is a \emph{connection}. Geometrically, a principal connection may be given as a \(G\)-equivariant horizontal distribution \(p\mapsto H_p\subset T_pP\) yielding a splitting \(T_pP = H_p \oplus V_p\), where \(V_p\) is the vertical subspace tangent to the group orbit through \(p\), together with the covariance condition
\be
(R_g)_*H_p = H_{p\cdot g}.
\ee
Equivalently, one can encode the same structure by a Lie-algebra valued one-form \(\omega\in\Omega^1(P,\mathfrak g)\) satisfying
\be\label{eq:equivariance}
(R_g)^*\omega = \mathrm{Ad}_{g^{-1}}\omega,
\qquad
\omega(\xi^\#)=\xi,
\ee
where \(\xi^\#\) is the fundamental vertical vector field generated by \(\xi\in\mathfrak g\).
The vertical bundle automorphisms act on connections, as expected, by pull-back \(\omega\mapsto \Phi^*\omega\). 

\begin{figure}[t]
\centering
\includegraphics[width=0.6\textwidth]{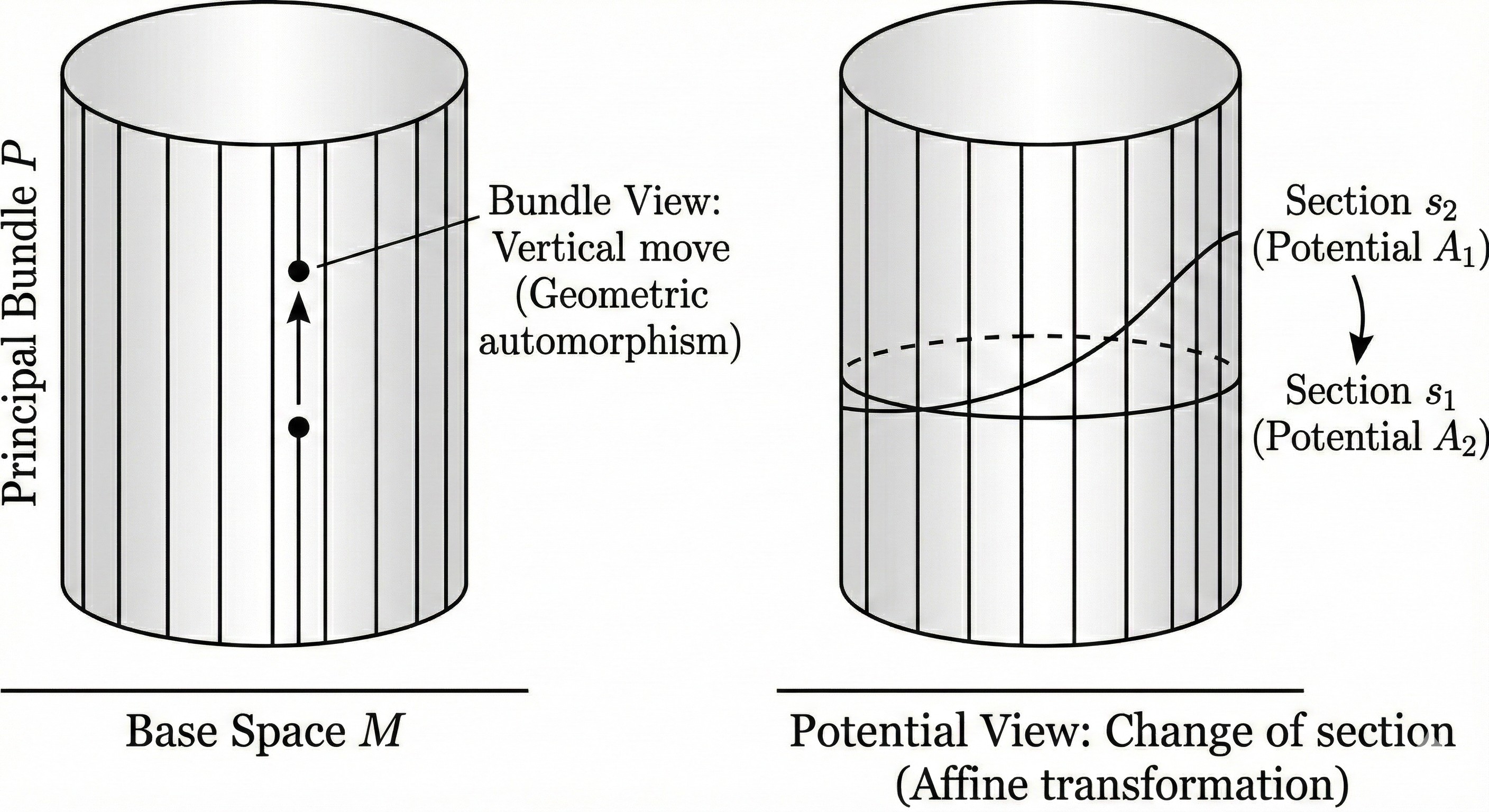}
\caption[Bundle view vs.\ potential view]{Left: in the principal-bundle formalism, a gauge transformation is a vertical move along the fibre---an automorphism of the bundle $P\to M$. Right: in the potential formalism, changing gauge corresponds to changing the section (the `cut' through the bundle); this is an affine transformation $A\mapsto A + d\chi$ of the local representative, not an automorphism of any natural background structure. }
\label{fig:aut_vs_gt}
\end{figure}

 Symmetry-related models are here related by automorphisms of the representational background. The difference between two connections \(\omega\) and \(\Phi^*\omega\) is, once again, a role-swap: not a change in gauge-invariant content, but a reassignment of which bundle points (or fibre elements) play which representational roles.

With respect to practice, one rarely attempts to extract invariant content from all the symmetry-related representatives of a connection form \(\omega\). For example, none of the textbooks surveyed here (cf. Section \ref{sec:survey}) include the transformation law for $\omega$ under a general $\Phi$ (the relation given by the first equation in \eqref{eq:equivariance} tells us only how the values of a single $\omega$ are related along a (vertical) direction of $P$.) The seminal \citep{kobayashivol1} mentions homomorphisms between \textit{different} principal fiber bundles (p.~53, 79, 80): \(F:(P', M', G')\rightarrow (P,M,G)\), and uses the tools gained only to study embeddings of one bundle into another (and reductions of the structure group). They establish a standard of identity between different backgrounds, but allow the automorphisms to remain implicit.\footnote{\cite{kobayashivol1}[Prop 6.1] describes the standard of identity as follows (disambiguating notation): let \(F:P'(M',G')\to P(M,G)\) be a bundle homomorphism with corresponding group homomorphism \(\varphi:G'\to G\), and let \(f:M'\to M\) be the induced diffeomorphism of bases. Let \(\varphi_*:\mathfrak{g}'\to \mathfrak{g}\) be the induced Lie-algebra homomorphism. Let \(\Gamma'\) be a connection in \(P'\), \(\omega'\) the connection form and \(\Omega'\) the curvature form of \(\Gamma'\). Then
\begin{enumerate}
\item[(a)] There is a unique connection \(\Gamma\) in \(P\) such that the horizontal subspaces of \(\Gamma'\) are mapped into horizontal subspaces of \(\Gamma\) by \(F\).
\item[(b)] If \(\omega\) and \(\Omega\) are the connection form and the curvature form of \(\Gamma\) respectively, then \(F^{*}\omega = \varphi_*\cdot \omega'\) and \(F^{*}\Omega = \varphi_*\cdot \Omega'\), where \(\varphi_*\cdot \omega'\) or \(\varphi_*\cdot \Omega'\) means the \(\mathfrak{g}\)-valued form on \(P'\) defined by
\[
(\varphi_*\cdot \omega')(X') = \varphi_*\big(\omega'(X')\big),
\qquad
(\varphi_*\cdot \Omega')(X',Y') = \varphi_*\big(\Omega'(X',Y')\big).
\]
\end{enumerate}\label{ftnt:PFB_iso}}

This is the same style of implicitness familiar from covariant GR: one fixes a connection \(\omega\) and studies curvature/holonomy, without having to keep watch over its entire gauge orbit. That is the whole story in the bundle case.

\subsection{Gauge potentials and the explicitness of gauge freedom}\label{subsec:gaugepotentials}

The moment one passes to local representatives, one introduces extra representational choices (coordinates, local trivialisation, gauge). In the gauge-potential formalism this forces the gauge freedom to be handled explicitly: one must keep track of how the local potentials transform under changes of section/trivialisation.

Fix a base manifold \(M\) and choose an open cover \(\{U_i\}\). In the gauge-potential formalism one starts from local Lie-algebra-valued one-forms (potentials) defined on the patches. One may regard the potentials as arising from choosing local sections (equivalently: a local trivialisation) of an underlying principal bundle; in the potential-first presentation that representational choice implies that changing section induces the familiar inhomogeneous gauge transformation on the local representatives. Accordingly, the gauge field on each patch is represented by a Lie-algebra-valued one-form
\be
A^{(i)} \;=\; A^{(i)}_\mu\,dx^\mu \in \Omega^1(U_i,\mathfrak g),
\ee
with compatibility conditions on overlaps.\footnote{Concretely: on each non-empty overlap \(U_i\cap U_j\) there exist smooth transition functions \(g_{ij}:U_i\cap U_j\to G\) such that \(g_{ji}=g_{ij}^{-1}\) and \(g_{ij}g_{jk}g_{ki}=\mathrm{Id}\) on triple overlaps. One then requires \(A^{(j)}\) to be related to \(A^{(i)}\) by a gauge transformation on \(U_i\cap U_j\). In the non-Abelian case this takes the familiar inhomogeneous form \(A^{(j)} = g_{ij}^{-1}A^{(i)}g_{ij} + g_{ij}^{-1}dg_{ij}\), while in the Abelian case it reduces to \(A^{(j)}=A^{(i)}+d\chi_{ij}\) for some functions \(\chi_{ij}\).}

In this setting, the natural candidates for ``automorphisms of the background'' are extremely meagre. If one takes the background to consist only of the patch \(U_i\) together with the vector space \(\Omega^1(U_i,\mathfrak g)\), then the obvious background automorphisms are diffeomorphisms of \(U_i\) and, at most, pointwise linear automorphisms of the Lie algebra \(\mathfrak g\).\footnote{There is a canonical group of Lie-algebra automorphisms \(\mathrm{Aut}(\mathfrak g)\), consisting of invertible linear maps \(\alpha:\mathfrak g\to\mathfrak g\) such that \(\alpha([X,Y])=[\alpha(X),\alpha(Y)]\). Such maps act pointwise on \(\mathfrak g\)-valued forms by \(A\mapsto \alpha\circ A\). In the Abelian case \(\mathfrak g=\mathbb R\), \(\mathrm{Aut}(\mathbb R)\cong\mathbb R^\times\) acts by constant rescalings \(A\mapsto \lambda A\). Crucially, these actions are linear and background-level: they do not depend on additional field-valued data. For this reason they do not implement the gauge redundancy of the potential formalism, which is affine (e.g.\ \(A\sim A+d\chi\) in electromagnetism) rather than linear.}

But this is not the gauge redundancy that matters for the potential formalism. Even in the Abelian case, the relevant equivalence relation is
\begin{equation}\label{eq:gt}
A^{(i)} \sim A^{(i)} + d\chi,
\end{equation}
an \emph{affine} shift of the potential by an exact form. This transformation is neither the pull-back by a diffeomorphism of \(U_i\) nor the pointwise action of a fixed Lie-algebra automorphism. It cannot arise as an automorphism of any background structure that the representation itself holds fixed. Accordingly, when the theory is presented purely in terms of local potentials on patches, gauge transformations invoke an additional equivalence relation on the space of potentials. This is just what it looks like when BRS fails in the gauge-potential formalism.

When we turn toward the usual textbook practice, we indeed find that it treats ``gauge invariance'' as a standing constraint. The gauge transformation property of the gauge potential foregrounds the entire development of the subject  and gauge invariance must be checked, enforced by gauge-fixing, or recovered indirectly via (traces of products of) curvatures, Wilson loops, etc. For a representative classical example, one often imposes the Lorenz gauge condition \(\partial_\mu A^\mu=0\) in Maxwell theory to decouple the equations and obtain a wave equation for the potential, \(\Box A_\nu = J_\nu\) (in suitable units). 
 One always remarks that the holonomy and curvature are gauge-invariant in the Abelian case; or, in the non-Abelian case, that traces of products of the curvature and of the holonomy are.  Examples of this kind are ubiquitous---too numerous to catalogue adequately here.

\paragraph{Summary (Gauge theory).}
The gauge-theory case fits the by now  expected pattern. In the 
principal-bundle formalism, gauge transformations are 
vertical bundle automorphisms---automorphisms of the 
background---and the symmetry stays implicit. In the 
potential formalism, gauge freedom is an additional 
equivalence relation on local representatives, and must be 
explicitly managed. The bundle formalism satisfies BRS; 
the potential formalism does not.

\section{Quantisation and Regional Subsystems}
\label{sec:quantum}

Sections~\ref{sec:GR}--\ref{sec:gauge} identified one clear route by which local symmetry becomes explicit: BRS fails, because the admissible background no longer renders the symmetry an automorphism. But BRS failure is not the whole story. Even when BRS holds, standard tasks in physics need to invoke symmetry explicitly.

The reason is that BRS concerns \emph{identity conditions} for individual models: it tells us when symmetry-related models may be treated as notational variants, so that one can proceed with a single representative without worry. Quantisation and subsystem analysis, by contrast, demand \emph{comparisons} between models or \emph{matches} between regions. Working merely ``up to isomorphism'' is not enough, because it leaves open how distinct representatives are to be related.

Throughout, the ontological stance stays the same: symmetry-related configurations still represent the same physical possibility. What changes is what we are trying to \emph{do}. The argument is not that quantisation `breaks' classical symmetries, nor is the issue peculiar to the Hamiltonian constraint or the problem of time: spatial diffeomorphisms raise exactly the same demand, even though they generate no analogous problem of time (cf.~\citep{ThebaultGryb_hole}). Gluing and boundary-matching, likewise, already require explicit structure in \emph{classical} field theory.

The globe example from Section~\ref{sec:sophistication} makes the point vivid. For a single globe one proceeds without the proviso `up to rotation'. But comparing two landmass distributions---Pangaea and the present---calls for a relational anchor that tells us what has `stayed in place'. Quotienting by rotations does not answer this; it tells us \emph{that} the two distributions differ globally, not \emph{how} they differ at each point.

That the need for cross-model identifications is a serious practical
problem---not merely a philosophical one---is plain from the literature. The problem of building diffeomorphism-invariant observables has long been recognised as central to canonical quantum gravity \citep{Isham_POT, Rovelli_book, HenneauxTeitelboim}, and has recently been taken up by other approaches. Here is how \citet{Harlow_cov} put it:
\begin{quote}
\ldots in gravity any local observable by itself will not be diffeomorphism-invariant, so we must dress it \ldots\ In practice such [invariant] observables are usually constructed by a `relational' approach: rather than saying we study an observable at some fixed coordinate location, we instead define its location relative to some other features of the state.
\end{quote}

I call any additional structure that supplies such cross-model or cross-regional identifications a \emph{representational scheme}. To supply a representational scheme the symmetry must be treated \emph{explicitly}, since one introduces machinery that picks out representatives and aligns gauge or diffeomorphism orbits so that comparison and gluing become well-defined.

\subsection{Quantisation: why one model is not enough}
\label{subsec:quantisation}

Three tasks force symmetry into view even when BRS holds. Let me spell out the demand. As argued in \citet{Rep_conv}, internal sophistication leaves two gaps:

\begin{enumerate}
\item[(A)] \emph{Individuation.} Sophistication commits the theory to a symmetry-invariant ontology, but does not hand us an explicit, symmetry-invariant \emph{description} of it. Given two models, we cannot in general check whether they are isomorphic by scanning the infinite-dimensional symmetry group. We need a complete set of invariants that can tell non-isomorphic models apart.

\item[(B)] \emph{Correspondence.} By identifying symmetry-related models, sophistication says nothing about how to match up objects in models that are \emph{not} isomorphic. Yet such matchings are indispensable for the quantum tasks at hand---superposition, path integration, local observables---and useful besides for expressing counterfactuals (`this region could have been more curved').
\end{enumerate}

For a single classical model of the whole universe, neither gap has teeth: one representative will do, and there is no call to compare it with others. Quantisation changes the situation, because every central construction---path integrals, constraint imposition, superposition---involves many models at once.

\paragraph{Path integrals and gauge-fixing.}
The path integral makes the point most directly. One integrates over field configurations weighted by the action:
\be
Z = \int \mathcal{D}\phi \, e^{iS[\phi]}.
\ee
If symmetry-related configurations are counted separately, the integral diverges. The Hessian $\delta^2 S / \delta\phi^2$ has zero modes along gauge directions---
\be
\frac{\delta^2 S}{\delta\phi^2} \cdot \delta_\xi \phi = 0
\ee
---so it is not invertible, and the propagator is undefined.

The standard remedy is gauge-fixing: restrict the integration to a submanifold $\mathcal{F}_{\mathrm{gf}} \subset \Phi$ that cuts each gauge orbit once, obtaining:
\be
Z = \int_{\mathcal{F}_{\mathrm{gf}}} \mathcal{D}\hat\phi \, \Delta_{\mathrm{FP}}[\hat\phi] \, e^{iS[\hat\phi]},
\ee
where $\Delta_{\mathrm{FP}}$ is the Faddeev--Popov determinant \citep{HenneauxTeitelboim}. Gauge-fixing is not merely a trick to cure a divergence. It introduces a representational scheme: the gauge-fixed variables $\hat\phi$ serve as (local) coordinates on $\Phi/\mathcal{G}$, so the integral over $\mathcal{F}_{\mathrm{gf}}$ is (locally) an integral over physical possibilities. Equivalently, gauge-fixing provides a local decomposition $\Phi\simeq \Phi/\mathcal{G}\times \mathcal{G}$, and the Faddeev--Popov determinant is the Jacobian relating the bare measure to the measure of that decomposition. Gauge-fixing thus amounts to choosing adapted coordinates on $\Phi/\mathcal{G}$ (Section~\ref{subsec:schemes}; see also \citep{Rep_conv}).

\paragraph{Dirac quantisation and local observables.}
In Dirac's canonical approach one promotes phase-space variables to operators and imposes the constraints as operator equations on the physical Hilbert space. For first-class constraints $C_a$ generating the gauge symmetry, physical states satisfy $\hat{C}_a |\Psi\rangle = 0$. For instance, in general relativity the momentum constraint generates spatial diffeomorphisms; a wavefunctional $\Psi[\gamma]$ on the space of spatial metrics is physical if and only if it is diffeomorphism-invariant. In the Schr\"odinger representation, with momentum operator $\hat\pi^{ij}(x) = -i\,\delta/\delta \gamma_{ij}(x)$, smearing the momentum constraint with a vector field $v^i$ on $\Sigma$ and integrating by parts yields
\be
-i\int_\Sigma d^3x\,\mathcal{L}_v \gamma_{ij}(x)\,\frac{\delta}{\delta \gamma_{ij}(x)}\Psi[\gamma]=:\delta_v\Psi[\gamma]=0.
\ee
The functional derivative of $\Psi$ along any orbit direction $\delta_v \gamma_{ij}=\mathcal{L}_v \gamma_{ij}$ must vanish: the constraint enforces invariance of the wavefunctional under spatial diffeomorphisms.
Global invariants satisfying this condition are easy to write down: any functional $\Psi[\gamma] = \int_{\Sigma} F[\gamma; x) \, d^3x$, with $F$ a scalar density, is automatically invariant.\footnote{I use the mixed dependence notation of DeWitt: square brackets for functional dependence, and round brackets for local.} But such quantities are non-local---they report on total curvature or total volume, not on what is happening \emph{here} rather than \emph{there}.

When we want \emph{local} gauge-invariant quantities, the gaps A and B above gain teeth. A naive observable like `the curvature at point $x$' is not diffeomorphism-invariant, because the label $x$ carries no invariant meaning. To get a local invariant, we must pin down $x$ relationally: `the curvature where the scalar field $\phi$ takes value $\phi_0$'. A representational scheme does exactly this: it anchors bare labels to invariant features of the configuration, and in doing so fixes what counts as `the same point' across different configurations---which is the cross-model identification that superposition and path integration require.

There is a good reason to think no alternative route exists. The gauge-invariant quantities one can write down \emph{without} a representational scheme are the non-local ones: integrals of scalar densities, holonomies, and other global invariants that never single out a point or a bounded region. Any gauge-invariant quantity that is genuinely local---that tells us about the physics \emph{here} rather than \emph{everywhere}---must say what `here' means in gauge-invariant terms, and saying that amounts to a dressing. The dressing evaluates an undressed quantity at a relationally defined point, and constructing that definition draws on the automorphism group. Locality and gauge-invariance, taken together, seem to require dressing; and dressing invokes symmetry explicitly.

That is the price of locality.

\paragraph{Superposition and cross-model comparison.}
Superposition is perhaps the case where the multi-model character is most visible---but the reason is slightly more subtle than it first appears. Superposition is addition in \emph{Hilbert space}, not in field space: the state $\alpha|\phi_1\rangle + \beta|\phi_2\rangle$ is a wavefunctional that assigns amplitudes to field configurations, not the field configuration $\alpha\phi_1 + \beta\phi_2$. Where, then, does the cross-model identification enter?

It enters because the Hilbert space is built \emph{over} the space of field configurations $\Phi$. A state is a wavefunctional $\Psi[\phi]$, and any local operation within that Hilbert space---computing an expectation value, restricting to a region, decomposing into subsystem factors---requires summing or integrating over, and thus comparison of, many configurations at once. For instance, `the expected curvature at $x$' is schematically
\be 
\langle \Psi | \hat{O}(x) | \Psi \rangle \;=\; \int \mathcal{D}\phi\; |\Psi[\phi]|^2 \; O(\phi;\, x),
\ee
and the label $x$ must pick out a coherent location across all the configurations in the integral. Without a representational scheme, `the point $x$' can arbitrarily shift meaning from one configuration to the next, even if the next is isomorphic to the first. 

To illustrate, suppose we want to define a tensor product state for a region $U$ and two field configurations $\phi, \phi'$, say $|\phi_U\rangle\otimes |\phi'_{U}\rangle$. On a gauge-invariant reading, there is no canonical way to hold fixed `the same $U$' across $\phi$ and $\phi'$. A scheme is needed to anchor what counts as the corresponding region across models. And it must be covariant: if $U$ is selected for $\phi$, then $g(U)$ must be selected for $\phi^g$. Otherwise the purported local construction would depend on the choice of gauge representative.

Why think this demand is hard to avoid? The physical configuration space $\Phi/\G$ is not a manifold in general (it has singularities at configurations with nontrivial stabilisers), but more importantly, there is no obvious way to define $L^2(\Phi/\G)$ as a working Hilbert space. So in practice one works in $L^2(\Phi)$ and imposes gauge-invariance as a constraint on states. But $L^2(\Phi)$ is built over the \emph{full} space of models, and any \textit{local} construction within it---expectation values at a point, regional decompositions, inner products that probe local data---requires a consistent identification of `where' across the configurations the wavefunctional is defined over. That identification is a representational scheme; I see no way to carry out local operations in the quantum theory that sidesteps it.


\subsection{Representational schemes: the geometry of comparison}
\label{subsec:schemes}

Section~\ref{subsec:quantisation} showed \emph{why} representational schemes are needed. I now describe \emph{what they are}, using the geometry of the space of models.

\paragraph{The space of models as a principal bundle.}
Fix a background structure $B$ (a smooth manifold, a principal bundle, background fields, or whatever the theory takes as given). The space of models $\Phi$ over $B$---the space of metrics, connections, or other dynamical fields---carries an action of the automorphism group $\G = \mathrm{Aut}(B)$. For concreteness, take gauge theory in the principal-bundle formalism of Section~\ref{subsec:PFB}: $B$ is the bundle $\pi:P\to M$, $\G = \mathrm{Aut}(P)$ is the group of bundle automorphisms (the vertical ones being gauge transformations), and $\Phi$ is the space of principal connections on $P$. The story for diffeomorphisms in general relativity is analogous.

\begin{figure}[t]
\centering
\includegraphics[width=0.5\textwidth]{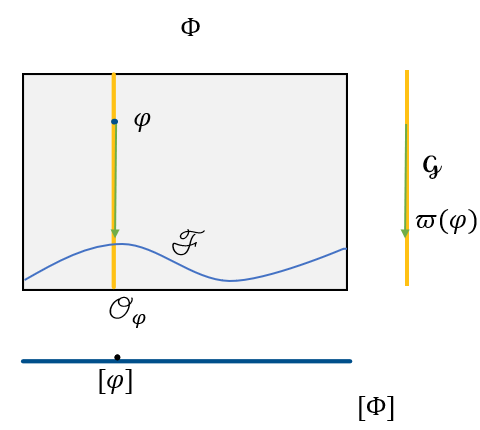}
\caption[The space of models]{The space of models $\Phi$ as a principal $\G$-bundle over the space of physical possibilities $[\Phi]$. Each gauge orbit $\mathcal{O}_\phi$ is a fibre over the equivalence class $[\phi]$. A gauge-fixing condition $F: \Phi \to V$ with $\dim V = \dim(\mathrm{Lie}(\G))$ defines a section $\mathscr{F} = F^{-1}(0)$ (shown in yellow) that intersects each orbit exactly once. The dressing $\varpi: \Phi \to \G$ sends each model $\phi$ to the unique group element $\varpi(\phi)$ such that $\phi^{\varpi(\phi)} \in \mathscr{F}$.}
\label{fig:model_bundle}
\end{figure}

When the action is free (which holds for generic configurations, away from those with nontrivial stabilisers), $\Phi$ has the structure of a principal $\G$-bundle over the quotient $[\Phi] := \Phi/\G$---the space of physical possibilities (see Figure~\ref{fig:model_bundle}). Both reduction and internal sophistication agree on this base space, with symmetry-related models identified. What neither gives us is a way to compare points in \emph{different} fibres: given two models $\phi_1$ and $\phi_2$ representing different physical possibilities, there is no canonical identification of the points of $B$ across them.

\paragraph{Gauge-fixing as a section.}
A representational scheme can be fixed (at least locally) by choosing a section $[\Phi] \to \Phi$: a rule that picks a unique representative from each equivalence class. In practice, one describes such a section as the zero-set of a gauge-fixing functional:
\[
\mathscr{F} = F^{-1}(0), \qquad F: \Phi \to V,
\]
where $V$ is a vector space of the right dimension to impose one condition per gauge direction. This ensures that $\mathscr{F}$ cuts each gauge orbit exactly once---at least locally, within a \emph{Gribov region}: a domain in field space where the gauge-fixing remains non-degenerate and single-valued. No single gauge-fixing provides a global section; the topology of $\Phi/\G$ generally obstructs any global choice.

Given such a gauge-fixing, every model $\phi$ within the Gribov region determines a unique group element $\varpi(\phi) \in \G$ such that the transformed model lies on the gauge-fixing surface: $F(\phi^{\varpi(\phi)}) = 0$. The map $\varpi: \Phi \to \G$ is called a \emph{dressing}. From the uniqueness of the solution, we get $\phi^{g\varpi(\phi^g)}=\phi^{\varpi(\phi)}$ so the dressing obeys a covariance equation:
\be\label{eq:varpi_cov}
\varpi(\phi^g) = g^{-1} \varpi(\phi).
\ee

\paragraph{Equilocality relations.}
To compare two models we need an \emph{equilocality relation}: a rule that says which point in one model corresponds to which point in the other. Formally, given models $(\phi_1, \phi_2)$, an equilocality relation is a map $\varepsilon: (\phi_1, \phi_2) \mapsto g_{12} \in \G$, where $g_{12}$ aligns $\phi_1$ with $\phi_2$: e.g. for diffeomorphisms on $M$ the point $x \in M$ in the description of $\phi_1$ corresponds to $g_{12}(x) \in M$ in the description of $\phi_2$ (cf. \citep{GomesButterfield_hole2}); \emph{mutatis mutandis} for gauge theory on a principal bundle,  for  points $p$ along a fiber and vertical automorphisms; or for gauge theory on a vector bundle $E$,  for vectors on each fiber and linear automorphisms of $E$ (cf. footnote \ref{ftnt:geom}). Different gauge-fixings yield different equilocality relations; there is no privileged choice.

But to be consistent, $\varepsilon$ must satisfy two constraints:
\begin{enumerate}
\item[(i)] \emph{Composition:} Comparing $\phi_1$ to $\phi_2$ and then $\phi_2$ to $\phi_3$ should give the same result as comparing $\phi_1$ directly to $\phi_3$:
\[
\varepsilon(\phi_1, \phi_3) = \varepsilon(\phi_1, \phi_2) \circ \varepsilon(\phi_2, \phi_3).
\]
This is a cocycle condition; as a special case, $\varepsilon(\phi, \phi) = \mathrm{id}$.

\item[(ii)] \emph{Covariance:} If $\phi_2 = \phi_1^g$ for some automorphism $g \in \G$ then the equilocality between them should just be that automorphism:
\be
\varepsilon(\phi, \phi^g) = g.
\ee
This avoids inconsistencies between different choices of isomorphic representatives.
\end{enumerate}

The dressing $\varpi$ induced by a gauge-fixing provides such a relation. Given two models $\phi_1$ and $\phi_2$, define
\[
\varepsilon(\phi_1, \phi_2) := \varpi(\phi_1) \circ \varpi(\phi_2)^{-1}.
\]
We can check both constraints directly.

\emph{Composition:}
\begin{align*}
\varepsilon(\phi_1, \phi_2) \circ \varepsilon(\phi_2, \phi_3) &= (\varpi(\phi_1) \circ \varpi(\phi_2)^{-1})(\varpi(\phi_2) \circ \varpi(\phi_3)^{-1}) \\
&= \varpi(\phi_1) \circ \varpi(\phi_3)^{-1} = \varepsilon(\phi_1, \phi_3).
\end{align*}

\emph{Covariance:} If $\phi_2 = \phi^g$, then using $\varpi(\phi^g) = g^{-1}\varpi(\phi)$:
\begin{align*}
\varepsilon(\phi, \phi^g) &= \varpi(\phi) \circ \varpi(\phi^g)^{-1} = \varpi(\phi) \circ (g^{-1}\varpi(\phi))^{-1} = g.
\end{align*}

\begin{figure}[t]
\centering
\includegraphics[width=0.5\textwidth]{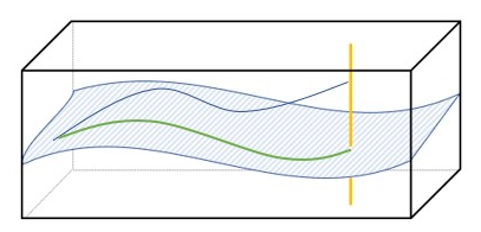}
\caption[Threading across models]{A one-parameter family of non-isomorphic models, $\phi_s$ (blue curve) projected to the gauge-fixing surface. The projected curve $\{\phi_s^{\varpi(\phi_s)}\}$ lies entirely in $\mathscr{F}$. The family of diffeomorphisms $\varpi(\phi_s)$ \emph{threads} spacetime points across non-isomorphic models: a point $x$ in $\phi_1$ corresponds to $\varepsilon(\phi_1, \phi_2)(x)$ in $\phi_2$.}
\label{fig:threading}
\end{figure}

\paragraph{From equilocality to dressed observables.}
A gauge-fixing also yields a complete set of gauge-invariant quantities---\emph{dressed
observables}---that solve the problem raised in Section~\ref{subsec:quantisation}:
they give the label $x$ a consistent, gauge-invariant meaning across configurations,
which is exactly what local operations in the quantum theory require.  They thereby describe each equivalence class of configurations in a gauge-invariant way.

For each gauge-fixing $F$, define the \emph{dressed model} $D(\phi) := \phi^{\varpi(\phi)}$. By construction, $D(\phi)$ lies on $\mathscr{F}$. The dressing covariance equation \eqref{eq:varpi_cov} ensures that $D$ is gauge-invariant:
\[
D(\phi^g) = \phi^{g\,\varpi(\phi^g)} = \phi^{g\,g^{-1}\varpi(\phi)} = \phi^{\varpi(\phi)} = D(\phi).
\]
Conversely, $D(\phi') = D(\phi)$ implies $\phi' = \phi^g$ for $g=\varpi(\phi)\varpi(\phi')^{-1} \in \G$. So $D(\phi)$ gives a complete, gauge-invariant description of the physical state.

Any (potentially gauge-variant) local quantity $\mathcal{O}(\phi; x)$ can now be dressed:
\[
\mathcal{O}^{\mathrm{dressed}}(\phi) := \mathcal{O}(D(\phi); x) = \mathcal{O}(\phi; \varpi(\phi)^{-1}(x)).
\]
This is relational: we evaluate the quantity not at the bare label $x$, but at the point that $x$ picks out once the gauge-fixing has aligned the models. Different gauge-fixings yield different dressed observables, but each set is equally complete. The choice of scheme determines which invariant quantities we foreground---it is not a discovery about which are `really' fundamental.


\subsection{Regional subsystems: composition and boundaries}
\label{subsec:regional}

Sections~\ref{subsec:quantisation} and \ref{subsec:schemes} concerned tasks that span many models: superposition, expectation values, and the construction of local observables. Regional subsystems raise a complementary demand that spans many \emph{regions}. The task is \emph{composition}: specify a region $R \subset M$ and its complement $\bar{R}$, describe the state on each, and glue them into a global state by matching data on $\partial R$.

In gauge theory and gravity, describing `fields on $R$ modulo symmetry' does not give us enough to carry out this gluing. The reason is not that the boundary breaks gauge symmetry---in Yang--Mills theory on a region with boundary, the local symmetry stays perfectly intact. Rather, it is the \emph{subsystem split} that forces symmetry into the open. Gauge redundancy that looks like mere relabelling from a global standpoint encodes, from the subsystem standpoint, genuine relational information: it tells us how the description on $R$ is oriented relative to the description on $\bar{R}$.

This is the core of \cite{RovelliGauge2013}'s argument that gauge redundancy exists precisely to enable subsystem coupling: it provides the `handles' by which independent regional descriptions can be joined. The recent edge-mode literature spells this out.\footnote{The origin of the mechanism was originally pointed out in \citep{DonnellyFreidel}; see \citep{GomesStudies} for a philosophical introduction and \citep{GomesRiello_new, Hohn_diffeo, carrozza2021, Ball2026} and references therein for an updated and thorough literature on the topic.\label{ftnt:refs}} Edge modes are the additional degrees of freedom that keep account of the relative gauge frame between $R$ and $\bar{R}$---exactly the relational information a representational scheme provides.

One vivid way to see the issue is through the failure of naive factorisation. In ordinary quantum mechanics, the Hilbert space of a composite system is the tensor product of the subsystem spaces: $\mathcal{H} = \mathcal{H}_R \otimes \mathcal{H}_{\bar{R}}$. In a gauge theory, this fails. The physical Hilbert space sits inside $\mathcal{H}_R \otimes \mathcal{H}_{\bar{R}}$ but does not fill it; the mismatch turns on the symmetry group acting at the boundary $\partial R$. Transformations that are `pure gauge' in a closed system can become boundary-nontrivial, so gluing requires data not encoded by bulk invariants alone. And indeed, the dynamical structures of a bounded subsystem---its Hamiltonian, symplectic structure, and variational principle---are not invariant under arbitrary symmetry transformations of the boundary data (cf footnote \ref{ftnt:refs}).

Once a representational scheme is fixed on one side of the boundary, the scheme on the other side can carry genuine physical content. Different global states can agree on the intrinsic state of $R$ and of $\bar{R}$ separately, yet differ in how the two regions are related---information that only the boundary data captures.\footnote{This is the many-to-one relation between global physical possibilities and collections of intrinsic subsystem states analysed in \citet{Gomes_new}. The resulting holism---global physical possibilities need not supervene on intrinsic subsystem states---is one concrete setting in which subsystem symmetries acquire direct empirical significance.} This is a form of holism: the global state is not determined by the intrinsic regional states, and the missing piece is relational.

Again, there are good reasons for thinking this is not a quirk of any one formalism. Any procedure that reconstructs global states from regional data---whether via edge modes, boundary conditions, or relational frames---must supply the information that fixes how the two regional descriptions line up. That information specifies a relative orientation in the symmetry group, and so brings the symmetry into the open. You cannot glue without saying \emph{how} the regions are aligned, and how they are aligned is a statement about the symmetry group.

One can supply the missing structure in different ways. One strategy enlarges the regional state space by boundary degrees of freedom so that symmetry acts canonically and gluing becomes well-defined. Another supplies an explicit relational identification via a representational scheme (a gauge-fixing, dressing, or relational frame), which need not coincide across $R$ and $\bar{R}$ but must control how the regions are stitched \citep{GomesStudies, GomesRiello_new}. Either way, regional composition requires matching data that goes beyond what bulk invariants provide; a representational scheme supplies exactly that.


\subsection{Summary}
\label{subsec:quantumsummary}

The limits of sophistication are not only \emph{background-sensitive} (Sections~\ref{sec:GR}--\ref{sec:gauge}) but also \emph{task-sensitive}. BRS tells us when symmetry can stay implicit for single-model work. But once we turn to tasks that are essentially multi-model---computing expectation values, building local observables, superposing configurations---or multi-region---composing subsystem descriptions, gluing across boundaries---new demands arise that working `up to isomorphism' does not meet. In each case the quotient $\Phi/\G$ falls short.

The machinery that meets these demands---representational 
schemes built from gauge-fixings and dressings---also reveals 
a connection between the two gaps identified at the start 
of the section. Individuation~(A) asks for a complete set 
of gauge-invariant quantities; correspondence~(B) asks for 
cross-model identifications. As Section~\ref{subsec:schemes} 
showed, a gauge-fixing that supplies equilocality relations 
simultaneously yields a complete set of dressed observables, 
and vice versa. Both gaps are closed by the same structure, 
and both routes work \emph{through} the background 
automorphisms; both bring symmetry into view.

How general is this? Each subsection offered reasons---grounded in how $\Phi/\G$ is put together---for thinking the demand is hard to dodge. This falls short of a proof that every conceivable formalism must employ a representational scheme under that name. But any adequate treatment of these tasks will need to introduce something that does the same work: supplying cross-model identifications or cross-region matching. And any such structure makes the symmetry group explicit.

Two caveats. First, representational schemes are not unique: different choices yield different but equally legitimate notions of `the same place' across models, and therefore different sets of dressed observables. Second, schemes are only local in field space: Gribov horizons prevent any single construction from working everywhere, so any particular scheme is at best a local chart on the space of physical possibilities. One might try to avoid both complications by restricting attention to global observables and declining to discuss local physics or subsystems. But that is a retreat from the tasks, not a way of accomplishing them without symmetry-handling. 

\section{Conclusion}\label{sec:conclusion}


The paper identified a fairly clean division. When a setting enriches or replaces the admissible background so that the symmetry no longer acts by automorphisms, BRS fails and symmetry must be handled explicitly---this is the \emph{background-sensitive} consideration (linearised GR, the initial value problem, the $3+1$ formalism, the gauge-potential formalism). Even when BRS holds, certain tasks---computing expectation values, building local observables, composing regional descriptions---require cross-model or cross-region identifications that the quotient $\Phi/\G$ does not provide, and supplying them brings the symmetry group into view---this is the \emph{task-sensitive} consideration. Together, the two yield a simple criterion: symmetry can stay implicit when BRS holds and the work at hand is exhausted by single-model description; when either condition fails, symmetry must be made explicit.


An account of sophistication that cannot recover this pattern is missing something. 
Many accounts begin from the question of what it is for a symmetry to be `merely  representational', and then argue over the mathematics and metaphysics that follow. But an account of sophistication is all the better if it can tell us where theorists will need to tackle symmetry, and where they can afford to ignore it. If a philosophical view predicts that practitioners should be doing something they manifestly are not, that view owes us an explanation.

The criterion developed here suggests further test cases. 
Asymptotic symmetries live at boundaries, where cross-region 
matching is unavoidable; the BMS literature bears this out. 
Dualities relate distinct theories---a form of cross-model 
comparison analogous to what Section~\ref{sec:quantum} 
studied within a single theory. Effective field theory 
matches descriptions across energy scales, which requires 
tracking how symmetries are realised at each level. Working 
through these cases would sharpen the picture considerably.

Symmetry is sometimes invisible in working physics, and 
sometimes unavoidable. Sophistication is most convincing 
when it can explain that difference.

\begingroup
\setlength{\emergencystretch}{8em}
\sloppy
\raggedright
\bibliographystyle{apacite}
\bibliography{references3}
\endgroup
\end{document}